\documentclass{article}

\usepackage{bioarxiv}
\usepackage[utf8]{inputenc} 
\usepackage[T1]{fontenc}    
\usepackage{booktabs}       
\usepackage{microtype}      
\usepackage{graphicx}       
\usepackage{amsmath}        
\usepackage{newtxtext}
\usepackage{newtxmath}
\usepackage{tabularx}       
\usepackage{setspace}
\usepackage{hyperref}       

\makeatletter
\renewcommand{\section}{%
  \@startsection{section}{1}{\z@}%
  {-2.0ex \@plus -0.5ex \@minus -0.2ex}%
  {1.5ex \@plus 0.3ex \@minus 0.2ex}%
  {\normalfont\large\bfseries\raggedright}%
}
\makeatother

\title{AI-Driven Scientific Computing Workflows: A Systems Review of Orchestration, Execution, Reproducibility and Provenance}

\author{
  Dr Jamie J. Alnasir\\
  Department of Computer Science\\
  Royal Holloway University of London\\
  Egham, England, UK \\
  \texttt{jamie.alnasir@rhul.ac.uk} \\
}

\begin{document}

\maketitle

\begin{abstract}
Artificial intelligence (AI) is increasingly embedded within scientific computing workflows that combine simulation, data processing, optimisation, visualisation and experimental or observational components. Learned models may serve as explicit workflow components, retain persistent state and, in adaptive settings, influence subsequent computation. Existing work has characterised scientific workflow management systems, dynamic and steered workflows, AI--HPC coupling motifs and the machine-learning lifecycle, although these areas are often discussed separately. This review brings them together from a systems perspective. We distinguish conventional scientific workflows, machine-learning pipelines, AI-coupled high-performance computing (HPC) workflows and broader automated research workflows, and propose a continuum describing the depth of AI participation from a computational stage to co-adaptive workflow control. The associated systems requirements are organised around five concerns: control and orchestration; compute and execution; data and model state; reproducibility and provenance; and governance and assurance. Representative systems and applications include AI-steered molecular simulation, drug and materials discovery, simulation--surrogate coupling and distributed self-driving laboratories. Workflow-level evaluation is considered in terms of scientific progress, execution cost, data movement, resource use, resilience and decision traceability. We conclude by identifying open problems in dynamic workflow representation, state-aware recovery, heterogeneous scheduling, interoperable data planes, model-mediated decision provenance and reproducible adaptive execution.
\end{abstract}

\noindent \textbf{Keywords:} Scientific Workflows $\cdot$ Artificial Intelligence $\cdot$ High-Performance Computing $\cdot$ Workflow Orchestration $\cdot$ Reproducibility $\cdot$ Provenance $\cdot$ Scientific Machine Learning

\section{Introduction}

Modern scientific computing increasingly relies on coordinated workflows that combine multiple computational and data-processing stages. A scientific result may depend on simulations, data transformations, model fitting, statistical analysis, visualisation, parameter exploration and post-processing, often executed across heterogeneous computing resources and storage systems. Scientific workflow management systems (WMSs) make these processes explicit, automatable and repeatable, and have become an important abstraction for large-scale science \cite{deelman2018future,deelman2015pegasus,suter2026terminology}. The workflow abstraction separates, to varying degrees, the scientific logic of a process from the mechanics of task execution, resource allocation, dependency resolution, retries, data movement and monitoring.

The increasing use of artificial intelligence (AI), machine learning (ML) and scientific machine learning within such workflows changes the systems problem in important ways. In the simplest case, an AI model is another executable stage: data are prepared, a trained model performs inference, and its output is consumed by a subsequent task. In more tightly integrated cases, AI participates in workflow control. Model outputs may rank candidates, terminate unproductive simulations, select new initial conditions, determine which fidelity of model should be invoked, trigger retraining, or alter an experimental or simulation campaign while it is running. In such settings, model outputs directly influence subsequent workflow execution.

Dynamic, iterative and steered scientific workflows predate the current wave of AI integration. Dynamic steering and runtime modification were established research topics before contemporary AI became widely embedded in scientific workflows \cite{oliveira2015steering,jain2015fireworks}. Recent community terminology likewise includes cycles, branches, adaptive behaviour and human-in-the-loop interaction alongside static directed acyclic graphs (DAGs) \cite{suter2026terminology}. The narrower systems question addressed here is: \emph{what changes when learned model outputs and model state become first-class participants in scientific workflow execution and control?}

Several adjacent literatures provide partial answers. Work on AI--HPC convergence has described dynamic and intelligent workflows \cite{ejarque2022hpcai,ferreiradasilva2024frontiers}, while Brewer et al.\ recently proposed six recurring execution motifs for online AI-coupled HPC workflows and analysed their middleware and performance implications \cite{brewer2025aicoupled}. The National Academies has used the broader term \emph{automated research workflows} for closed-loop research processes that combine computation, AI, experiments and observations \cite{nasem2022arw}. Scientific-ML work has addressed lifecycle management and provenance \cite{souza2022provenance}, while the wider workflow community has developed interoperable approaches to run provenance and FAIR workflow descriptions \cite{leo2024rocrate,wilkinson2025fair}. These contributions are complementary. The present review treats them as parts of a common systems problem spanning control, execution, state, reproducibility and provenance.

We provide a systems-level synthesis of these adjacent strands of work. Earlier practical guidance has addressed HPC usage and the design of AI-driven HPC workflows \cite{alnasir2021hpc,alnasir2026tips}; the present article develops an analytical framework for their systems-level integration. The contribution has four parts. We first define the scope of \emph{AI-driven scientific computing workflows} and distinguish them from ordinary ML pipelines, conventional scientific workflows and broader automated research workflows. We then introduce a continuum for the depth of AI participation in computation, decisions and state. Five interacting systems concerns --- control and orchestration; compute and execution; data and model state; reproducibility and provenance; and governance and assurance --- provide the basis for analysing the resulting requirements. Finally, representative systems and applications are compared through this lens and used to identify unresolved research questions. The resulting perspective complements motif-oriented AI--HPC surveys. Its purpose is analytical synthesis; system ranking and benchmark evaluation lie outside scope.

\section{Scope, Terminology and Review Approach}

\subsection{Scientific workflows and workflow management systems}

The term \emph{scientific workflow} has broadened substantially over the past two decades. Deelman et al.\ emphasised the importance of automated workflow management for future extreme-scale science \cite{deelman2018future}; more recently, Suter et al.\ proposed a community-developed terminology organised around five axes: workflow structure and characteristics, composition, orchestration, data management and metadata capture \cite{suter2026terminology}. Their definition explicitly accommodates dynamic, adaptive and interactive workflows, including cycles and human participation. We adopt that broad definition here.

A workflow is not synonymous with a DAG, even though DAG-based representations remain common. DAGs are attractive because dependencies are explicit and scheduling is tractable, but some scientific processes require runtime branching, iterative refinement, event-driven behaviour, long-lived services or externally triggered tasks. WMSs such as Pegasus, Nextflow, Snakemake, FireWorks and Parsl occupy different points in this design space \cite{deelman2015pegasus,ditommaso2017nextflow,moelder2021snakemake,jain2015fireworks,babuji2019parsl}. Their abstractions, execution models and intended domains differ, and this diversity is one reason that no single WMS has become a universal solution \cite{suter2026terminology}.

\subsection{ML pipelines, AI-coupled HPC and automated research workflows}

An ML pipeline is commonly organised around data ingestion and transformation, model training, validation, deployment and inference. Such a pipeline may itself be represented as a workflow, but its scientific objective is often the production or application of a model. By contrast, an AI-driven scientific computing workflow embeds one or more learned models within a larger scientific process whose objective may be simulation, discovery, optimisation, inference about a physical system, or generation of a scientific dataset.

Throughout this review, \emph{AI} refers primarily to data-driven learned models, including scientific-ML models and foundation-model or agentic components. Rule-based, heuristic and optimisation mechanisms may also appear in adaptive workflows, but they fall within the present framework only where learned-model outputs or model state participate in workflow execution, control or scientific progression.

Adaptive workflows make the distinction especially clear. A conventional ML pipeline may end after a model is selected and deployed. An AI-driven scientific workflow may instead execute a loop such as
\[
\text{simulation} \rightarrow \text{analysis} \rightarrow \text{AI inference}
\rightarrow \text{selection} \rightarrow \text{new simulation},
\]
possibly with periodic model retraining. DeepDriveMD is a representative example: simulation data are used to train models, inference identifies scientifically informative conformations, and those predictions influence subsequent molecular-dynamics simulations \cite{brace2022deepdrivemd}. Colmena similarly supports event-driven, AI-guided steering of ensemble calculations \cite{ward2025colmena}.

Brewer et al.\ distinguish AI-in-HPC, AI-out-HPC and AI-about-HPC coupling and derive six execution motifs --- dynamic orchestration, multistage pipelines, inverse design, digital replicas, distributed models and adaptive training \cite{brewer2025aicoupled}. That taxonomy is particularly useful for describing recurring online AI--HPC interaction patterns. Our concern is somewhat broader: we include offline and online AI components, scientific ML lifecycle state, reproducibility and provenance, and workflows that may span HPC, cloud, edge or experimental infrastructure.

Automated research workflows (ARWs) are broader still. The National Academies describes ARWs as research processes that integrate computation, laboratory automation and AI across the design, execution and analysis of experiments, simulations and observations \cite{nasem2022arw}. Our use of \emph{AI-driven scientific computing workflow} deliberately centres the computational systems problem. Instrument-connected and self-driving laboratories are included when they illuminate orchestration, state or provenance requirements, but physical laboratory automation is not treated as the paper's primary domain.

\subsection{Review approach}

This article uses a structured narrative systems-review approach. Relevant literature was identified iteratively through targeted searches of scholarly indexes, publisher repositories and preprint servers, together with citation chaining from established workflow, AI--HPC and scientific machine-learning literature. The review does not follow a formal systematic-review or meta-analysis protocol. Searches were updated through August 2026. Search vocabulary centred on three broad areas: scientific workflows and workflow management; AI--HPC and scientific machine learning; and lifecycle concerns including provenance, model state, reproducibility, data management and workflow evaluation. Established terminology papers, reviews and widely used workflow systems were used as anchor points, with additional sources incorporated where they provided relevant systems, methodological or application-level evidence.

Selection was purposive and systems-oriented. Sources were selected for their relevance to the systems questions addressed in this review. This included work on workflow terminology and infrastructure, AI-mediated control and state, workflow evaluation and benchmarking, provenance, reproducibility, governance and assurance. Peer-reviewed archival publications were preferred where available. Preprints were retained for rapidly developing topics when no definitive peer-reviewed version was identified and are labelled as such in the bibliography. Routine domain applications in which machine learning was used only as an isolated predictor, without material workflow or systems implications, were outside scope.

The systems and applications discussed below were selected to span different workflow abstractions, coupling patterns and scientific settings. The aim was to cover contrasting systems characteristics, not to construct an exhaustive catalogue or ranking. The four-level AI-participation continuum and the five-concern systems framework were developed through conceptual synthesis of recurring requirements and distinctions in the reviewed literature, and were considered alongside established workflow terminology and adjacent AI--HPC classifications \cite{suter2026terminology,brewer2025aicoupled}. They are used here as analytical constructs for comparison and are not presented as empirically validated or mutually exclusive taxonomies.

The review has corresponding limitations. Its narrative methodology does not support claims of exhaustive coverage, literature frequency or publication trends, and relevant systems may have been omitted. The emphasis on papers that expose coupling, control and state also under-represents domain studies in which AI is scientifically important but architecturally conventional. The purpose is to synthesise systems requirements and trade-offs; measuring the prevalence of particular workflow technologies lies outside that scope.

\section{AI Participation in Workflow Control and State}

AI-driven workflows can be distinguished by \emph{how deeply a learned model participates in workflow control and state}. We propose a four-level continuum (Table~\ref{tab:continuum}). The levels are analytical positions and need not be mutually exclusive; a single application may move between them during its lifecycle.

We operationalise the continuum using two distinctions. First, does a learned model merely compute a result, or does its output alter workflow execution? Second, if it alters execution, is that influence a bounded decision within an otherwise specified control structure, a recurrent modification of campaign-level execution state, or part of a loop in which the model state itself is updated from workflow-generated evidence? Level A denotes a predetermined AI task. Level B denotes model-mediated selection among bounded downstream alternatives. Level C denotes recurrent modification of campaign-level state --- for example task generation, cancellation, reprioritisation, parameter changes or termination --- during or between execution iterations. Level D adds co-evolution of model state and scientific workflow state. The B--C boundary is thus defined by the scope and persistence of model-mediated control, whereas the C--D boundary is defined by whether model state is itself updated as part of the scientific feedback loop. These criteria do not require synchronous or low-latency coupling: offline versus online execution remains an orthogonal dimension.

\begin{figure}[ht]
\centering
\includegraphics[width=\textwidth]{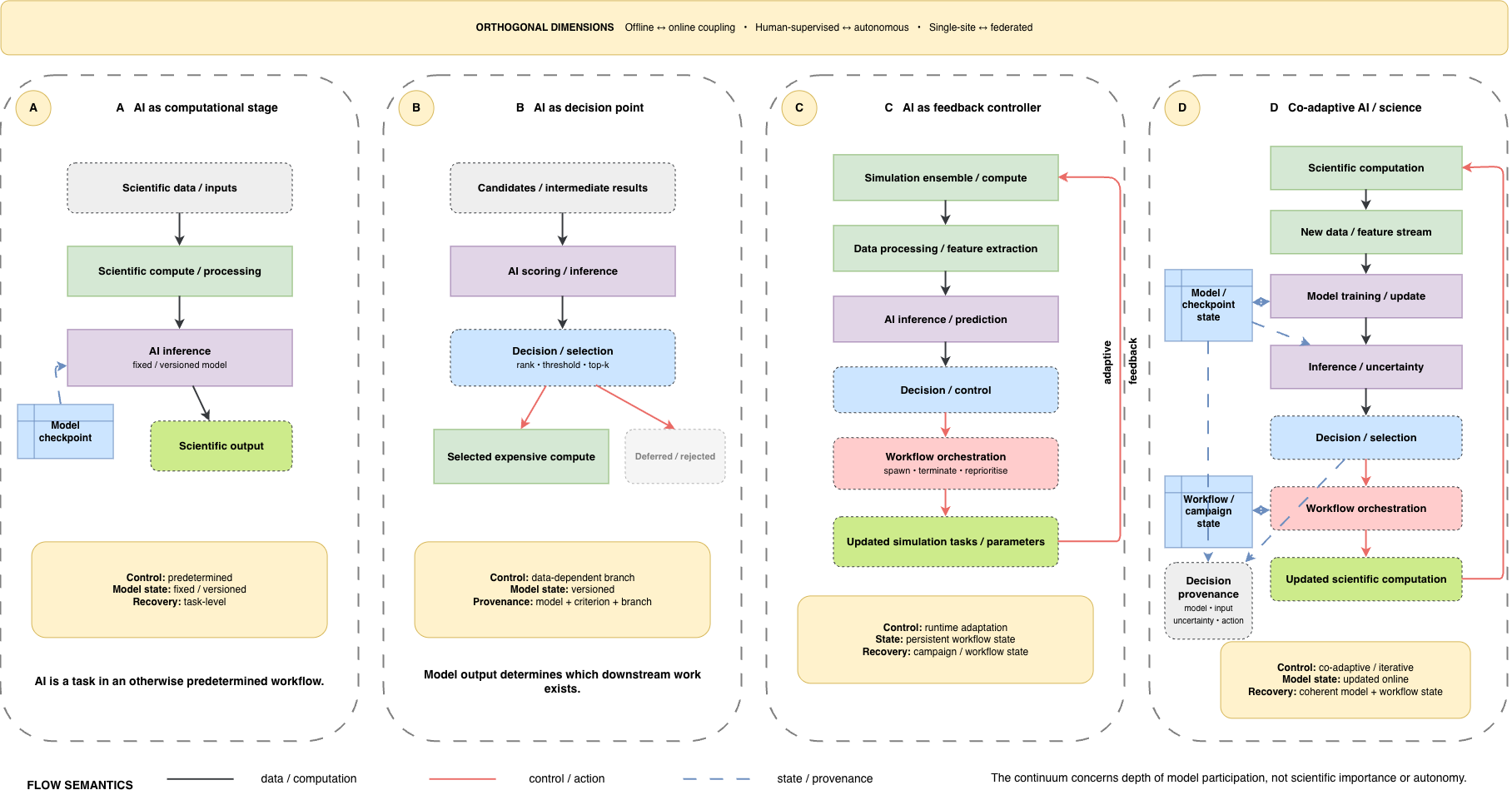}
\caption{Progressive participation of AI in scientific workflow control and state. Four representative modes are shown: (A) AI as a computational stage within a predetermined workflow; (B) AI as a decision point that determines which downstream work proceeds; (C) AI as a feedback controller that modifies a running computational campaign; and (D) co-adaptive AI/science, in which model state and scientific workflow state evolve together. The progression concerns the participation of learned model outputs and model state in workflow execution. Autonomy, scientific importance, offline versus online coupling, human supervision and federation are orthogonal characteristics that may vary at any level.}
\label{fig:ai-participation-continuum}
\end{figure}

\begin{table}[ht]
\centering
\caption{A continuum of AI participation in scientific workflow execution.}
\label{tab:continuum}
\begin{tabularx}{\textwidth}{p{0.19\textwidth}X X}
\toprule
\textbf{Level} & \textbf{Role of AI} & \textbf{Principal systems implication} \\
\midrule
AI as computational stage &
Training or inference is a predetermined task in the workflow. &
Conventional dependency management may suffice; model artefacts and specialised resources must nevertheless be represented explicitly. \\
AI as decision point &
Model output selects, filters or ranks bounded downstream alternatives within an otherwise specified control structure. &
Control flow becomes data-dependent; provenance must capture the model, input state and criterion responsible for the branch. \\
AI as feedback controller &
Model output recurrently modifies campaign-level execution state, such as task generation, cancellation, reprioritisation, parameters or termination. &
Persistent workflow state, dynamic task management and adaptive resource allocation become important; low-latency data exchange is required only when demanded by the application. \\
Co-adaptive AI/science &
Workflow-generated evidence updates model state while the updated model influences subsequent scientific computation. &
Workflow and model state become mutually dependent; coherent checkpointing, recovery, uncertainty handling and decision traceability become first-class concerns. \\
\bottomrule
\end{tabularx}
\end{table}

\subsection{AI as a computational stage}

At the first level, the workflow topology is substantially known in advance. A model may classify an image, score molecular candidates, estimate a surrogate quantity or transform a dataset. The task may require a GPU, a model checkpoint and a specific runtime, but it need not alter the execution semantics of the workflow. This pattern is close to conventional dataflow and is well handled by systems that can express explicit dependencies and resource requirements.

Even here, AI introduces lifecycle state that ordinary file-oriented pipelines may under-specify. The relevant executable is not just source code: reproducibility may depend on model architecture, weights, preprocessing transforms, tokenisers, feature definitions, training data versions, random seeds and hardware-dependent numerical behaviour. Consequently, treating the model as an opaque executable weakens reproducibility even if the surrounding task graph is deterministic.

\subsection{AI as a decision point}

At the second level, model outputs determine which bounded downstream alternative proceeds. Examples include ranking compounds before expensive simulation, rejecting low-value parameter combinations, selecting regions of interest in images, or choosing among alternative solvers or fidelity levels. Level B selects within an otherwise specified control structure. Persistent modification of campaign-level execution state characterises Level C. The workflow may remain largely stage based at Level B, but its realised execution path now depends on a learned model.

This distinction matters for provenance. A record that task B followed task A is not sufficient to explain why B was chosen instead of C. At minimum, a reproducible record must identify the model version and input state, the decision rule or threshold, and the resulting branch. Where the model output is probabilistic, confidence or uncertainty estimates may also be part of the decision context.

\subsection{AI as a feedback controller}

At the third level, model-mediated decisions recurrently modify campaign-level execution state. Dynamic steering of scientific workflows is not unique to AI \cite{oliveira2015steering}; the difference is that control decisions are now derived partly from learned representations or predictions. DeepDriveMD demonstrates this pattern in molecular simulation, while Colmena provides a general steering abstraction in which agents respond to task completion and other events \cite{brace2022deepdrivemd,ward2025colmena}. Tasks may be created, cancelled or reprioritised, parameters may change, and stopping criteria may be altered as evidence accumulates. Such control can occur online within a continuously active campaign or between asynchronous iterations; synchrony is not what defines this level.

At this level, the systems requirements extend beyond bounded branch selection. The orchestrator must maintain persistent workflow state, accept late-bound information, support dynamic task management, and avoid turning the steering component into a serial bottleneck. Where decisions are latency sensitive, file-mediated coupling may be inadequate and in-memory or service-based exchange may be preferable; where they are not, asynchronous or batch-mediated feedback can implement the same control semantics.

\subsection{Co-adaptive AI and scientific computation}

The most tightly integrated case is co-adaptive execution, in which model state and scientific workflow state evolve together. This distinguishes Level D from a feedback controller that repeatedly acts using a fixed or externally versioned model. Simulations generate new training data; the model is updated as part of the same scientific process; the new model changes sampling, control or optimisation; and the resulting computation generates further data. Inverse design, active learning, adaptive surrogate modelling and some digital-replica workflows fall in this class \cite{brewer2025aicoupled}. Such workflows challenge the assumption that a job can be restarted from a small collection of output files. Recovery may require a consistent snapshot spanning simulation state, model weights, optimiser state, data-selection history and orchestration state.

The degree of human involvement is orthogonal to this continuum. A human may approve model-selected candidates, intervene when uncertainty is high, or alter the scientific objective. Conversely, a relatively simple model may operate autonomously. Online coupling, cross-facility federation and physical instrumentation are also represented as orthogonal characteristics of a workflow.

\subsection{Relationship to AI--HPC coupling motifs}

The continuum in Figure~\ref{fig:ai-participation-continuum} is deliberately not a replacement for the execution motifs proposed by Brewer et al.\ \cite{brewer2025aicoupled}. Their motifs describe recurrent functional and dynamic coupling patterns between AI and HPC components; our continuum describes the \emph{depth of model participation in workflow control and state}. The two views are orthogonal and can be used together. For example, a multistage pipeline can contain AI only as a predetermined stage, or it can use model outputs to decide whether downstream high-fidelity stages execute. Likewise, a digital replica can be a passive predictive component or part of a bidirectional feedback loop.

A second adjacent perspective is the agentic-workflow framework of Shin et al., which organises the evolution of scientific workflows along dimensions of intelligence and composition, from static single workflows towards intelligent multi-agent or ``swarm'' configurations \cite{shin2025agentic}. That framework addresses the organisation and autonomy of workflow intelligence. The present continuum instead asks how learned-model outputs and model state participate in scientific execution and persistent workflow state. An agentic workflow may therefore occupy different positions on this continuum depending on whether its agents merely prepare tasks, make bounded decisions, steer campaign state or update their own models as part of the scientific loop.

Table~\ref{tab:brewer-map} illustrates the relationship. The mapping is intentionally non-exclusive: the same motif may span more than one level depending on the application implementation.

\begin{table}[ht]
\centering
\caption{Relationship between Brewer et al.'s AI--HPC execution motifs \cite{brewer2025aicoupled} and the AI-participation continuum used in this review.}
\label{tab:brewer-map}
\begin{tabularx}{\textwidth}{p{0.24\textwidth}p{0.33\textwidth}X}
\toprule
\textbf{Execution motif} & \textbf{Typical position on the continuum} & \textbf{Systems emphasis} \\
\midrule
Dynamic orchestration & Feedback controller; co-adaptive in online-learning variants & Runtime task generation/cancellation, control latency, steering state \\
Multistage pipeline & Computational stage to decision point & Data-dependent filtering, stage imbalance, intermediate-state transfer \\
Inverse design & Feedback controller to co-adaptive & Iterative optimisation, convergence state, repeated simulation--model exchange \\
Digital replica & Computational stage to feedback controller & Low-latency inference, bidirectional data/control flow, model update \\
Distributed models & Orthogonal to the continuum & Federation, network placement, cross-site data movement and identity \\
Adaptive training & Decision point to co-adaptive & Dynamic training, checkpoint/model state, simulation-guided data generation \\
\bottomrule
\end{tabularx}
\end{table}

This distinction is central to the scope of the present review. Brewer et al.\ focus on online AI--HPC execution motifs, middleware and performance \cite{brewer2025aicoupled}; here, those execution patterns are embedded within a wider lifecycle analysis that also includes offline model stages, persistent model/workflow state, reproducibility, provenance, governance and assurance.

\section{A Systems Framework for AI-Driven Scientific Workflows}

\begin{figure}[ht]
\centering
\includegraphics[width=\textwidth]{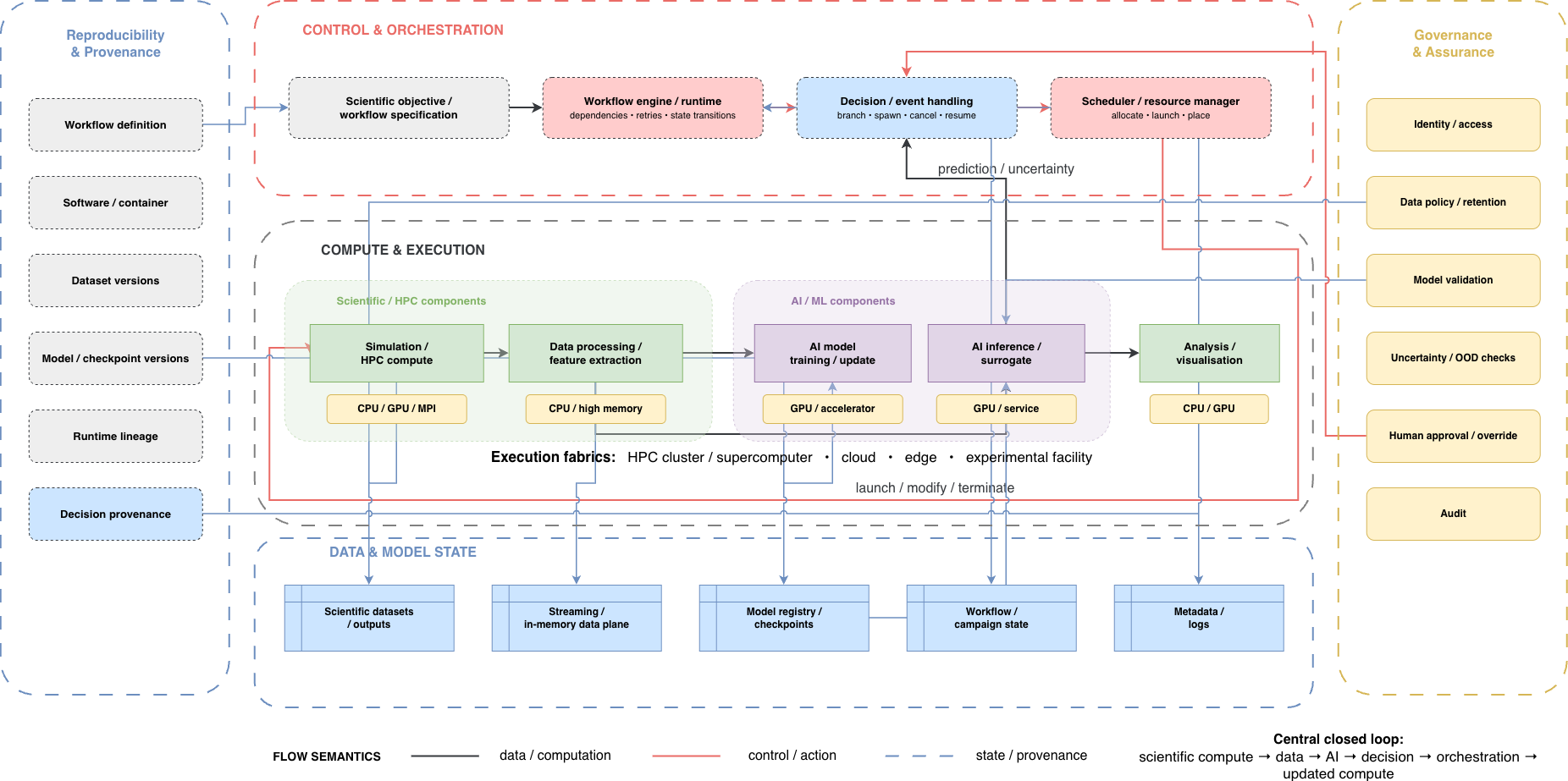}
\caption{Systems architecture of an AI-driven scientific computing workflow. Control and orchestration coordinate scientific objectives, runtime decisions and resource allocation; heterogeneous scientific/HPC and AI/ML components execute across the compute layer; and scientific data, model artefacts and workflow state are maintained in the data and model-state layer. Reproducibility and provenance, and governance and assurance, act as cross-cutting concerns across the architecture. Black arrows denote data or computation flow, red arrows denote control or action flow, and blue arrows denote state or provenance relationships. The central adaptive loop allows model outputs and uncertainty estimates to influence subsequent workflow execution.}
\label{fig:systems-framework}
\end{figure}

The continuum above describes \emph{how AI participates}; Figure~\ref{fig:systems-framework} provides a complementary view of \emph{what systems capabilities are affected}. We organise its requirements around five interacting concerns. Each concern predates current AI integration, but learned model outputs and model state alter the requirements placed on it and the interactions between layers. The framework is aligned with established workflow-system terminology \cite{suter2026terminology}.

\subsection{Control and orchestration}

This concern determines what executes, in what order, under which conditions, and how runtime events alter future execution. It encompasses dependency resolution, task generation, branching, loops, event handling, retries, cancellation, human intervention and interaction between the WMS and lower-level schedulers.

\subsection{Compute and execution}

This concern covers where and how tasks run: CPUs, GPUs and other accelerators; distributed memory and shared memory parallelism; batch schedulers; job arrays; long-lived workers; service endpoints; clouds; edge devices; and cross-facility execution. AI training, inference and conventional simulation frequently have different computational profiles, so a workflow may be heterogeneous even when all tasks run within one HPC facility.

\subsection{Data and model state}

Scientific workflows have always been data dependent, but AI increases the diversity and persistence of relevant state. In addition to conventional inputs and outputs, a workflow may carry checkpoints, feature stores, embeddings, optimiser state, replay buffers, candidate queues, data-selection histories and model versions. Where decisions depend on intermediate state, the distinction between ``data'' and ``control state'' becomes blurred.

\subsection{Reproducibility and provenance}

Reproducibility requires more than preservation of the task graph. Software environments, model artefacts, workflow parameters, datasets, random seeds, hardware context and runtime decisions may all affect the result. Provenance provides the trace linking these entities and activities. Existing work on scientific-ML provenance and Workflow Run RO-Crate provides a strong foundation \cite{souza2022provenance,leo2024rocrate}; AI-driven adaptive workflows add the requirement to record model-mediated decision context.

\subsection{Governance and assurance}

Where model outputs can steer expensive simulations, physical experiments or analyses of sensitive data, model validity and uncertainty become operational concerns. Governance includes access control, data retention, auditability and handling of derived sensitive artefacts. Assurance includes validation, uncertainty quantification, monitoring for out-of-distribution inputs and appropriate human oversight. Recommendations such as REFORMS highlight reproducibility, validity and generalisability concerns in ML-based science \cite{kapoor2024reforms}; scientific-ML uncertainty quantification provides methods for representing predictive uncertainty and avoiding deterministic interpretation of probabilistic model outputs \cite{psaros2023uq}.

The five concerns are coupled. A model-mediated decision in the control plane consumes model and data state, executes on a particular resource, and produces provenance that may need to satisfy governance requirements. Analysing one concern in isolation can simply displace a bottleneck to another layer.

\section{Control and Orchestration}

\subsection{Models as explicit workflow components}

Hiding AI logic inside a large script or monolithic executable can simplify initial implementation, but may weaken observability and make it harder to reason about resource use, failure modes, model versions and reproducibility. Where the surrounding workflow abstraction permits, representing models as explicit components with declared inputs, outputs, resource requirements and artefact dependencies can improve inspectability, scheduling and provenance. This is consistent with the broader workflow practice of separating scientific logic from execution mechanics \cite{ditommaso2017nextflow,moelder2021snakemake}.

It is likewise useful to distinguish training, validation and inference because they have different failure semantics and scaling behaviour. Training may be long-running, distributed and checkpoint intensive; inference may be latency sensitive or high throughput; validation may be a gating operation whose result changes subsequent execution. Combining all three phases in a single opaque task can make both scheduling and provenance less precise.

\subsection{From static dependency graphs to event-driven control}

Static DAGs remain suitable for many AI-enabled workflows. Difficulties arise when the graph itself changes as new evidence becomes available. FireWorks was designed for high-throughput applications requiring dynamic workflow modification \cite{jain2015fireworks}; the older literature on workflow steering likewise considered runtime intervention and adaptation \cite{oliveira2015steering}. AI-driven steering is a modern instance of a broader dynamic-workflow problem, distinguished by the fact that a model may generate or mediate the control signal.

Colmena illustrates an event-oriented approach. Its steering logic can respond to completed tasks, invoke inference, retrain models and submit new tasks, allowing the scientific strategy to evolve while resources remain active \cite{ward2025colmena}. Brewer et al.\ identify similar requirements across dynamic orchestration, adaptive training and other AI--HPC motifs \cite{brewer2025aicoupled}. A central research question is how to express such adaptation without hiding it inside application code, because hidden control logic is harder to inspect, reproduce and port.

\subsection{Workflow engines and runtime systems}

General-purpose and domain-oriented WMSs expose different orchestration abstractions. Pegasus focuses strongly on planning and execution of scientific workflows across distributed resources \cite{deelman2015pegasus}. Nextflow and Snakemake emphasise reproducible data pipelines and portability across execution back ends \cite{ditommaso2017nextflow,moelder2021snakemake}. Parsl exposes Python-based dataflow and dynamic task execution across heterogeneous resources \cite{babuji2019parsl}. Ray provides a distributed execution substrate oriented towards AI applications and stateful actors \cite{moritz2018ray}. FireWorks supports dynamic modification of workflows during execution \cite{jain2015fireworks}. These systems should not be reduced to a simple feature checklist: their programming models and assumptions differ.

ExaWorks takes a compositional approach in which multiple workflow technologies can interoperate across exascale use cases \cite{alsaadi2021exaworks}. The approach is particularly relevant to AI-driven science, where a single end-to-end process may combine a batch-oriented scientific WMS, a distributed training runtime, an inference service and a facility-specific scheduler.

\subsection{AI-assisted workflow management}

More recent work places AI directly at the workflow-management layer. Thareja et al.\ separate workflow intent, design and implementation through a structured specification stage, use an LLM-based debugging agent, and integrate Pegasus with a Model Context Protocol layer for workflow submission, monitoring and control \cite{thareja2026aiassisted}. Their system addresses workflow construction and runtime management; this role is distinct from a model that steers the underlying scientific computation.

\section{Compute and Execution}

\subsection{Heterogeneous resource requirements}

AI-driven workflows frequently alternate between computational regimes. Data cleaning and feature extraction may be CPU- and I/O-intensive, while model training may require multiple GPUs and collective communication. Inference may run on CPUs, GPUs or specialised accelerators depending on latency and batch size, whereas simulation may require MPI across many CPU or GPU nodes. Reserving all resources for the duration of a monolithic workflow can leave expensive accelerators idle, while releasing resources between every short task can introduce queueing and launch overhead.

Resource-aware decomposition becomes important in such heterogeneous workloads. Tasks should expose their actual requirements, and orchestration should permit independently scheduled stages when the loss of locality is justified. Efficient progress of the scientific campaign is the relevant objective; uniformly high device utilisation may be neither necessary nor desirable.

\subsection{Workflow-level parallelism}

Scientific workflows expose several forms of parallelism: parallel tasks, data parallelism, ensemble parallelism, parameter sweeps and model-training parallelism. These forms interact. A thousand independent simulations can provide abundant task parallelism, but coupling them to one centralised inference process may create a serial bottleneck. Conversely, distributing the model may introduce communication overhead larger than the inference cost.

Parsl, Ray and scheduler-native job arrays all provide mechanisms for exploiting high task concurrency \cite{babuji2019parsl,moritz2018ray}. In HPC settings, the choice between many scheduler submissions and a pilot or worker-pool model is important because scheduler overhead and queue policies can dominate fine-grained tasks. AI workflows intensify this issue when inference or lightweight preprocessing tasks are much shorter than simulations or training jobs.

\subsection{Coupling strength and communication}

The coupling between simulation and AI can be tight, semi-tight or loose. Tight coupling may place inference in the same process or node as a simulation, reducing latency but increasing integration complexity and potentially constraining resource placement. Looser coupling through an inference service improves modularity and allows independent scaling, but introduces network and serialisation overhead. Brewer et al.\ show that coupling strategy is itself a performance variable, not merely an implementation detail \cite{brewer2025aicoupled}.

SmartSim demonstrates a practical in-memory approach for connecting numerical simulations and ML models across language boundaries, using an in-memory database to avoid repeated file-based exchange \cite{partee2022smartsim}. DeepDriveMD and Colmena similarly show that adaptive-campaign performance depends on the end-to-end interaction between simulation, analysis, training, inference and orchestration \cite{brace2022deepdrivemd,ward2025colmena}.

Coupling intensity and AI-participation depth are distinct. A model can be invoked at high frequency and with very low latency while remaining a predetermined computational stage if its outputs do not alter workflow control. Conversely, an infrequent model invocation can constitute deeper participation when it changes task generation, termination, resource allocation or persistent campaign state. Coupling motifs and the participation continuum describe different properties of the workflow.

\subsection{Throughput and time-to-scientific-result}

Traditional HPC performance analysis often focuses on time-to-solution for a fixed calculation, strong or weak scaling, FLOP/s and communication efficiency. These remain important, but an AI-guided workflow may deliberately spend additional computation on training or inference to reduce the number of expensive simulations required. In that case, a slower component can improve the overall scientific campaign if it enables better prioritisation.

Accordingly, workflow-level measures such as time-to-scientific-result and useful outcomes per unit time can be more informative than isolated kernel speed. DeepDriveMD reports acceleration in terms of simulated time and conformational-state coverage, which captures the campaign objective more directly than neural-network execution speed \cite{brace2022deepdrivemd}. The relevant performance objective is thus tied to scientific utility and must be defined carefully for each application.

\section{Data and Model State}

\subsection{Data gravity and locality}

Placing computation near large datasets to reduce repeated data movement is a long-standing principle in data-intensive computing and remains relevant on HPC systems. AI can intensify data gravity because training repeatedly revisits datasets, checkpoints can be large, and intermediate representations may be consumed by multiple downstream components. Cross-facility workflows add wide-area transfer and authentication costs.

Model placement is partly a data-placement decision. A remote inference service may be attractive operationally but inappropriate if each request requires transferring large scientific arrays. Conversely, centralising a relatively small model near large simulation outputs may be preferable to moving those outputs to a remote accelerator.

\subsection{The HPC--AI I/O mismatch}

HPC simulation and AI training often have different I/O access patterns. Large simulations commonly favour high-bandwidth writes of large arrays through parallel I/O libraries and formats such as HDF5 and ADIOS2 \cite{folk2011hdf5,godoy2020adios2}. Training workloads may instead issue repeated, randomised reads of many samples. Brewer et al.\ identify this mismatch as a central data-path challenge for AI-coupled workflows \cite{brewer2025aicoupled}.

The consequence is that simply placing AI and simulation on the same parallel filesystem does not guarantee efficient coupling. Caching, data staging, format conversion, sharding and in-memory exchange may be necessary. Older coupled-workflow systems such as DataSpaces already demonstrated the value of decoupling producers and consumers through memory-oriented data services \cite{docan2010dataspaces}; AI makes these techniques newly important because data-production and model-consumption rates may differ substantially.

\subsection{Persistent model and workflow state}

Model weights are only one part of AI state. Reproducible restart of training may require optimiser state, learning-rate schedule, random-number generator state, training-data order and distributed-training metadata. Adaptive workflows add a second category: scientific decision state. A candidate queue, acquisition function, set of previously sampled configurations, uncertainty estimates and stopping criteria may determine future execution even if all raw simulation outputs are retained.

Checkpointing should be defined at the level of the \emph{workflow state that is necessary for semantically correct continuation}, not only at the level of each executable. In co-adaptive workflows, inconsistent checkpoints can be problematic: restoring a simulation ensemble from one point and a model from another may produce a valid program execution that no longer corresponds to the original scientific trajectory.

\subsection{Streaming, in situ processing and federation}

When decisions must be made in near real time, file-based boundaries can become an avoidable source of latency. ADIOS2 supports multiple data transport modes for high-performance applications \cite{godoy2020adios2}, while SmartSim provides a model for in-memory simulation--AI exchange \cite{partee2022smartsim}. In both cases, data transport is treated as an explicit systems choice instead of assuming that every task boundary requires a persistent file.

Federated workflows extend this issue across sites. Instruments, edge devices, cloud services and HPC systems may all participate in one workflow. Brewer et al.'s distributed-model motif and the National Academies' ARW framing both point towards increasingly distributed research infrastructures \cite{brewer2025aicoupled,nasem2022arw}. Latency, data sovereignty, identity management and partial failure become part of workflow design.

\section{Reproducibility, Provenance, Governance and Assurance}

\subsection{Reproducible environments are necessary but insufficient}

Containerisation has become a central technique for packaging scientific software environments. Singularity, now continued as Apptainer, was designed specifically to provide container mobility in HPC settings \cite{kurtzer2017singularity}. Workflow systems can bind containers to individual tasks, helping separate software dependencies from the underlying cluster \cite{ditommaso2017nextflow}.

Containers do not, however, reproduce an AI-driven workflow by themselves. A container image does not identify the exact training data, model weights, random seed, workflow parameters, external model registry state, runtime branch decisions or hardware context. Nor does it guarantee bitwise reproducibility of accelerator-based computation. Environment capture is one layer of reproducibility, not a substitute for provenance.

\subsection{Experiment tracking and scientific provenance}

ML experiment-tracking systems such as MLflow record model parameters, metrics and artefacts \cite{zaharia2018mlflow}. Scientific workflows require an additional layer: the relationships between scientific inputs, transformations, simulations, models and outputs. Souza et al.\ propose a provenance view spanning the scientific-ML lifecycle and show how W3C PROV-compatible representations can integrate domain and ML information \cite{souza2022provenance}.

Workflow Run RO-Crate addresses interoperability across WMSs by packaging workflow-run provenance together with associated inputs, outputs and code, and has been implemented across multiple workflow systems \cite{leo2024rocrate}. The FAIR workflows recommendations likewise treat workflows as research objects that should be findable, accessible, interoperable and reusable, with identifiers, metadata and explicit relationships to components and data \cite{wilkinson2025fair}. AI-driven workflow provenance can build on established standards while adding model- and decision-specific context.

\subsection{Provenance of model-mediated decisions}

Adaptive AI workflows add an important provenance question: \emph{why did the workflow take this path?} Conventional execution provenance can record that task A generated an output consumed by task B. For a model-mediated branch, scientific interpretation may also require the model identifier and version, the exact input presented to the model, preprocessing state, inference parameters, uncertainty or confidence estimate, the decision rule, and the control action generated from that output.

Dynamic workflow provenance predates contemporary AI. Learned models add a decision-producing artefact whose behaviour may be stochastic, data dependent and versioned independently from the workflow definition. We argue that decision provenance should be recorded explicitly when AI outputs affect scientific execution. Recent work on PROV-AGENT demonstrates one concrete direction: it extends W3C PROV to capture agent-centric metadata such as prompts, responses and decisions and relates these records to broader workflow context across heterogeneous environments \cite{souza2025provagent}. Model- or agent-mediated decisions can thus be represented within an end-to-end provenance model instead of being relegated to an isolated log.

PROV-AGENT is nevertheless aimed principally at foundation-model agents, whereas the requirement identified here is broader. Numerical surrogates, classifiers and active-learning models may lack prompts or conversational traces. Their decisions can still depend on checkpoints, preprocessing state, uncertainty estimates and application-specific control policies. A general decision-provenance layer needs to relate the model invocation and its state to the scientific input, decision rule, resulting control action and downstream consequence. Such records are important for debugging, audit, reproducibility and post hoc assessment of whether a scientifically important branch was produced by a robust model or by an artefact of model drift or uncertainty.

\subsection{Governance and assurance}

Governance becomes especially important when workflows handle clinical, genomic, proprietary or otherwise sensitive data. Model artefacts themselves may encode information derived from restricted datasets, and intermediate representations can fall outside simplistic policies that govern only raw input and final output. Access control, retention, audit and cross-site transfer policy need to apply to the workflow's derived state.

Assurance is related but distinct. REFORMS identifies recurring validity, reproducibility and generalisability failures in ML-based science and provides reporting recommendations intended to improve scientific reliability \cite{kapoor2024reforms}. In adaptive workflows, such model-quality questions have operational consequences: an overconfident or out-of-distribution prediction may redirect expensive computation or a physical experiment. Scientific-ML uncertainty quantification offers methods for representing predictive uncertainty \cite{psaros2023uq}, but workflow assurance must connect those estimates to explicit control policies --- for example, rejecting low-confidence decisions, escalating them to a human, triggering additional simulation or constraining the set of permitted actions. The assurance boundary is therefore wider than model validation alone: it includes the conditions under which a model is authorised to affect workflow state and the mechanisms for audit, override and recovery when that authority is exercised.

\section{Systems and Scientific Realisations}

\subsection{Workflow and runtime systems}

Table~\ref{tab:systems} summarises representative systems by the role they play in AI-driven scientific computing. The purpose is not to rank them; several are complementary and operate at different layers.

\begin{table}[ht]
\centering
\caption{Representative systems relevant to AI-driven scientific computing workflows. Entries describe concrete execution capabilities; the table does not rank the systems, and several operate at different layers that may be composed.}
\label{tab:systems}
\small
\begin{tabularx}{\textwidth}{p{0.11\textwidth}p{0.19\textwidth}Xp{0.23\textwidth}}
\toprule
\textbf{System} & \textbf{Primary abstraction} & \textbf{Runtime adaptation/control mechanism} & \textbf{Execution scope and relevance} \\
\midrule
Pegasus &
planned scientific workflows &
plan-based execution with runtime monitoring, retries and recovery; workflow structure is primarily specified before execution &
distributed/HPC; planning, data movement and provenance \\

Nextflow &
dataflow processes and channels &
process activation follows channel availability; composition supports data-dependent execution &
local/HPC/cloud; portable, containerised pipelines \\

Snakemake &
rule/dependency workflows &
checkpoints and input functions support data-dependent expansion after intermediate results become available &
local/cluster/cloud; reproducible rule-based analysis \\

FireWorks &
dynamic task graph &
running tasks can add, detour or defuse downstream workflow elements &
HPC/cluster; runtime task-graph mutation \\

Parsl &
Python dataflow tasks &
programmatic task submission using futures and dependency-driven execution over configurable executors &
HPC/cloud/distributed; high-throughput heterogeneous execution \\

Ray &
tasks and stateful actors &
dynamic task and actor creation with long-lived distributed state and services &
distributed clusters; AI and stateful service execution \\

Colmena &
event-driven steering agents &
steering logic reacts to completed work and can submit, prioritise or terminate subsequent tasks &
HPC ensembles; AI-guided campaign steering \\

SmartSim &
simulation--AI coupling &
application-controlled orchestration with in-memory data exchange and model serving alongside simulations &
HPC/distributed; low-latency simulation--model interaction \\
\bottomrule
\end{tabularx}
\end{table}

Pegasus, Nextflow and Snakemake demonstrate established approaches to scientific workflow description, dependency management and portable execution \cite{deelman2015pegasus,ditommaso2017nextflow,moelder2021snakemake}. FireWorks, Parsl and Ray expose programmatic mechanisms for runtime task creation, dependency-driven execution or stateful distributed components \cite{jain2015fireworks,babuji2019parsl,moritz2018ray}. Colmena and SmartSim address AI-integrated scientific computing more directly through event-driven steering and simulation--model coupling \cite{ward2025colmena,partee2022smartsim}. ExaWorks is notable because it treats composability among workflow technologies as a design objective \cite{alsaadi2021exaworks}, a pragmatic response to the observation that orchestration, distributed training, in-memory data services and HPC scheduling may be better provided by specialised components.

\subsection{AI-steered biomolecular simulation}

DeepDriveMD provides one of the clearest examples of AI participating in the control loop of a large-scale scientific workflow. Ensembles of molecular-dynamics simulations generate trajectory data; ML models learn representations of the conformational space; inference identifies promising states; and subsequent simulations are launched from those states \cite{brace2022deepdrivemd}. The application raises all five concerns of our framework: dynamic orchestration, heterogeneous simulation and training resources, high-volume state movement, the need to identify model and simulation state consistently, and the assurance question of whether ML steering could bias sampling towards artefacts.

The reported acceleration is also instructive methodologically. ML-guided sampling seeks to reach scientifically relevant conformational coverage with less aggregate simulation; shortening the execution time of an individual MD timestep is not the target. Effective workflow performance can diverge from component performance.

\subsection{Multistage virtual screening and drug discovery}

Virtual screening illustrates AI as both a stage and a decision mechanism inside a heterogeneous scientific campaign. IMPECCABLE (Integrated Modeling PipelinE for COVID Cure by Assessing Better LEads) integrates ML prediction with multiple physics-based stages of increasing cost and fidelity \cite{alsaadi2021impeccable}. Its architecture is important for this review because the model is not the scientific endpoint: AI is used to reduce and prioritise a vast candidate space before more expensive docking and binding free-energy calculations. The workflow couples algorithmic selectivity to resource allocation.

IMPECCABLE also provides a useful performance vocabulary. Al Saadi et al.\ distinguish \emph{throughput} (ligands processed per unit time), \emph{scientific performance} (effective ligands sampled per unit time) and \emph{peak performance} in FLOP/s \cite{alsaadi2021impeccable}. This distinction directly supports a systems-level view of AI-driven workflows. A method can reduce raw throughput yet improve scientific performance if it filters candidates more effectively; conversely, high accelerator utilisation is of limited value if it accelerates uninformative work. The case motivates the evaluation criteria in Section~10.

From the perspective of our framework, such multistage discovery workflows stress several parts of the systems stack. Orchestration must accommodate stages with dissimilar runtimes and resource needs, while execution spans CPU, GPU and ensemble workloads. Large candidate sets are progressively filtered, and provenance must retain the model and threshold responsible for eliminating a candidate because both can affect the scientific interpretation of the final shortlist.

\subsection{Materials discovery and distributed autonomous experimentation}

Materials discovery demonstrates how AI-driven computation can extend beyond an HPC facility while retaining the same systems concerns. Pyzer-Knapp et al.\ describe discovery cycles that combine AI, HPC simulation and robotic experimentation, with the components operating as a heterogeneous, iterative workflow \cite{pyzerknapp2022materials}. Such workflows naturally cross the boundary between computational orchestration and automated research infrastructure.

Bai et al.\ make the provenance and federation problem particularly explicit in a distributed self-driving laboratory linking robotic facilities in Cambridge and Singapore \cite{bai2024sdl}. A dynamic knowledge graph represents data and material flows, autonomous agents execute portions of the design--make--test--analyse cycle, and provenance is recorded as the knowledge graph evolves. This example lies near the outer boundary of our computational focus, but it is useful precisely because model-mediated decisions become physical actions. At that point, federation, semantic state, provenance, uncertainty and governance are not optional metadata concerns; they determine whether the scientific process can be audited and safely reproduced.

\subsection{Simulation--surrogate coupling}

Surrogate models can replace or approximate expensive components of a simulation, provide initial conditions, estimate closure terms, or act as digital replicas. SmartSim demonstrates online ML inference within a large-scale numerical ocean simulation and provides cross-language interfaces through an in-memory data layer \cite{partee2022smartsim}. This class of workflow highlights a different systems problem from ensemble steering: latency and data locality can dominate because the model may be invoked repeatedly within or alongside simulation timesteps.

The architecture chosen for inference --- embedded, node-local service, distributed service or remote endpoint --- changes both performance and failure behaviour. A loosely coupled service is easier to update and scale independently, but the network becomes part of the simulation's critical path. Tight coupling lowers latency but can reduce modularity and portability.

\subsection{Cross-case synthesis through the participation continuum}

Taken together, these examples show why AI--workflow coupling and AI-participation depth should be analysed separately. DeepDriveMD lies close to the co-adaptive end of the continuum because workflow-generated simulation data are used to update learned representations that subsequently influence new simulation work. The multistage screening pattern exemplified by IMPECCABLE is closer to a model-mediated decision point: learned predictions reduce or prioritise a candidate set before more expensive downstream calculations. SmartSim, by contrast, demonstrates that tight online simulation--model coupling does not by itself imply deep participation in workflow control; a repeatedly invoked surrogate can remain a predetermined computational stage when its invocation and downstream use are fixed by the surrounding application. Distributed self-driving laboratories can occupy feedback-control or co-adaptive positions depending on whether learned model state is itself updated within the experimental loop.

The comparison also reinforces the orthogonality of the additional dimensions introduced in Section~3. A workflow may be tightly or loosely coupled, human-supervised or substantially autonomous, and single-site or federated at any of several participation levels. The continuum identifies which parts of the systems stack become scientifically consequential as learned outputs and model state exert greater influence over execution and persistent workflow state; it is not intended as a ranking of application sophistication.

\section{Comparative Synthesis and Evaluation Criteria}

\subsection{Why component benchmarks are insufficient}

Benchmarking AI models and HPC kernels remains useful, but it cannot by itself characterise an adaptive AI-driven workflow. A training benchmark measures how rapidly a model reaches a target quality; a simulation benchmark measures solver performance; neither captures the delays, data movement, idle resources and control decisions created by coupling them. IMPECCABLE provides a useful concrete example: its authors distinguish raw throughput, scientific performance and peak FLOP/s because each answers a different question about a heterogeneous discovery campaign \cite{alsaadi2021impeccable}. This separation is preferable to treating the fastest component as synonymous with the fastest route to a scientific result.

MLPerf HPC provides an important benchmark suite for scientific ML training on HPC systems \cite{farrell2021mlperfhpc}. At the workflow level, WfCommons provides infrastructure for analysing real workflow instances and constructing representative synthetic workflows for workflow research and benchmarking \cite{coleman2022wfcommons}. Brewer et al.\ argue that AI-coupled workflows require additional workflow-specific benchmarks and coupling-aware analysis \cite{brewer2025aicoupled}. We agree, but this review does not propose or execute a new benchmark. Instead, we identify evaluation dimensions that can be used to interpret published systems work.

\subsection{Evaluation dimensions}

At least six dimensions are relevant.

\textbf{Time-to-scientific-result} measures elapsed time to a scientifically meaningful endpoint, not merely completion of one task. The endpoint must be stated explicitly, for example a target uncertainty, conformational coverage, validated candidate set or number of accepted discoveries.

\textbf{Throughput} measures useful scientific work completed per unit time. It is particularly relevant for high-throughput screening, ensemble simulation and hyperparameter exploration.

\textbf{Resource efficiency} includes accelerator utilisation, CPU efficiency, memory pressure and queueing overhead. Utilisation is useful insofar as it supports scientific progress; severe under-utilisation may nevertheless indicate poor decomposition or coupling.

\textbf{Data movement and I/O} measure bytes transferred, storage traffic, metadata operations and time spent staging or transforming data. These costs can dominate when AI changes access patterns.

\textbf{Resilience and recovery} measure the cost of failures and the ability to restart without corrupting workflow semantics. In adaptive workflows, recovery should be evaluated at the level of coherent workflow and model state.

\textbf{Traceability} measures whether the run can be reconstructed sufficiently to explain scientific outputs and runtime decisions. This includes both conventional provenance and model-mediated decision context.

Energy and carbon cost may also be relevant, particularly where AI is added to reduce total simulation cost. Such comparisons should be made end-to-end: an AI component that consumes additional energy can still reduce the footprint of the overall campaign if it avoids much more expensive computation.

\subsection{Comparing workflow classes}

The expected importance of these dimensions rises with AI participation. A predetermined AI stage can often be evaluated using conventional workflow metrics plus model quality. A feedback-controlled workflow additionally requires control latency, dynamic scheduling overhead and decision provenance. A co-adaptive workflow further requires measures of state consistency, retraining cost and the interaction between model uncertainty and scientific progress.

Future benchmark design should pair the task graph with a \emph{control contract}: what events may alter execution, what state is persistent, what constitutes a scientifically useful outcome, and what failure/recovery semantics are expected. Without such a contract, two systems may appear to execute the same ``workflow'' while providing materially different scientific behaviour.

\section{Open Challenges and Research Agenda}

\subsection{Workflow representations beyond static task graphs}

Modern workflow terminology already accommodates dynamic structures \cite{suter2026terminology}, yet many production systems and user practices remain easiest to reason about when dependencies are known before execution. AI-guided workflows require representations that make dynamic branches, loops and model-driven task generation explicit without sacrificing portability or inspectability. Embedding all adaptive logic in arbitrary Python solves expressiveness at the cost of analysability; restricting workflows to static DAGs solves analysability at the cost of expressiveness. Finding useful intermediate representations remains an open design problem.

\subsection{State-aware fault tolerance and recovery}

Task retry is insufficient when correctness depends on correlated state across simulation, model and orchestrator. Future systems need checkpoint semantics that can identify a consistent recovery point across multiple components, potentially with different checkpoint frequencies and storage costs. This is particularly challenging for federated workflows in which components fail independently.

\subsection{Adaptive heterogeneous scheduling}

Resource requests are currently often specified per task. Co-adaptive workflows can change the mix of simulation, training and inference dynamically. Schedulers and WMSs need mechanisms that respond to changing demand without excessive queue delay or resource fragmentation. Pilot-job and elastic-worker approaches offer partial solutions, but cross-resource optimisation remains difficult, especially when accelerator types are heterogeneous.

\subsection{Interoperable data planes}

The coexistence of HPC data formats, ML dataloaders, streaming systems, in-memory stores and object storage creates conversion and movement overhead. AI-coupled workflows would benefit from higher-level data abstractions that preserve semantics while allowing the runtime to choose transport and storage. ADIOS2, DataSpaces and SmartSim demonstrate important pieces of this space \cite{godoy2020adios2,docan2010dataspaces,partee2022smartsim}, but the systems reviewed here do not provide a single data plane spanning every workflow pattern and facility.

\subsection{Decision provenance}

Workflow provenance standards are increasingly mature \cite{leo2024rocrate,wilkinson2025fair}, and PROV-AGENT shows how agent-specific prompts, responses and decisions can be incorporated into W3C-PROV-compatible workflow provenance \cite{souza2025provagent}. Model-mediated scientific control nevertheless raises broader questions. How should a workflow record a decision derived from a stochastic numerical model? Which aspects of the inference environment are necessary to reproduce the branch? How should uncertainty and the application-level decision policy be represented? Should a retrained model constitute a new workflow component version, a new state of the same component, or both? Answering these questions in interoperable provenance models would improve both reproducibility and scientific audit.

\subsection{Reproducibility of adaptive execution}

Strict replay of an adaptive workflow may be impossible or scientifically undesirable when models, random sampling and distributed execution are nondeterministic. Reproducibility may therefore need to be defined at several levels: exact execution replay, equivalent control decisions, statistically equivalent scientific outcomes, or reproducible reconstruction of the evidence that justified each decision. The appropriate level is domain dependent and should be reported explicitly.

\subsection{Governance of derived model artefacts}

Scientific governance frameworks often distinguish raw data from published outputs. AI workflows introduce many intermediate artefacts --- checkpoints, embeddings, synthetic data and fine-tuned models --- that may retain information from restricted datasets. Policies and workflow systems need to represent these artefacts explicitly, propagate access constraints and support audit across sites.

\subsection{Foundation models and agentic orchestration}

Foundation models and AI agents may serve as scientific predictors or as orchestration components that generate candidate workflows, interpret failures, propose parameter changes or coordinate tools. Shin et al.\ describe this trajectory through dimensions of workflow intelligence and composition \cite{shin2025agentic}. Thareja et al.\ demonstrate an AI-assisted workflow-management approach that separates specification from implementation, uses an LLM-based debugging agent and integrates Pegasus with a Model Context Protocol layer for execution and interaction \cite{thareja2026aiassisted}. In these approaches, learned systems participate directly in workflow management as well as domain computation.

Higher-level planning and tool use make the same systems requirements more stringent. Generated plans must be validated before execution, agent actions must be bounded by explicit policies, provenance must distinguish proposed plans from executed activities, and high-impact operations may require human approval or override. The intelligence/composition perspective of agentic workflows is complementary to the participation continuum proposed here: an agent may assist specification without controlling scientific execution, or it may progress towards feedback control and co-adaptation. Agentic orchestration extends the responsibilities of workflow management and assurance.

\section{Conclusions}

AI-driven scientific computing workflows allow learned models to participate at different depths in computation, persistent state and workflow control. Dynamic and adaptive workflows predate contemporary AI. The distinguishing feature considered here is the use of learned model outputs and model state as operational inputs to workflow decisions and scientific progression.

We have analysed that problem across five interacting concerns: control and orchestration; compute and execution; data and model state; reproducibility and provenance; and governance and assurance. Existing WMSs, AI runtimes and data services provide many of the required capabilities, although the surveyed systems distribute them across multiple abstractions. Applications such as DeepDriveMD, Colmena, SmartSim and distributed self-driving laboratories show that the behaviour of the coupled system can influence scientific value independently of the performance of any single solver or model.

Future evaluation should consider workflow-level outcomes together with scientific quality and uncertainty. Relevant measures include time-to-scientific-result, throughput, data movement, utilisation, resilience and traceability. Reproducibility also extends beyond software and data to model versions, persistent state and the decisions that changed execution. As learned models take a larger role in workflow control, scientific computing infrastructure will need stronger support for state, interoperability and provenance.

\section*{Declarations}

\textbf{Author contributions.} J.J.A. conceived the review, conducted the literature synthesis, developed the conceptual framework and figures, and wrote and revised the manuscript.

\textbf{Data and code availability.} No primary research datasets or new software were generated as part of this review. The article synthesises information from the cited literature.

%
%

\end{document}